# User recommendation in reciprocal and bipartite social networks
## --a case study of online dating

Kang Zhao, Xi Wang, Mo Yu, and Bo Gao

**Abstract**

Many social networks in our daily life are bipartite networks built on reciprocity. How can we recommend users/friends to a user, so that the user is interested in and attractive to recommended users? In this research, we propose a new collaborative filtering model to improve user recommendations in reciprocal and bipartite social networks. The model considers a user's "taste" in picking others and "attractiveness" in being picked by others. A case study of an online dating network shows that the new model has good performance in recommending both initial and reciprocal contacts.

# 1. Introduction

With billions of users, online social networking websites not only change how people search, receive, and spread information, but also affect the way people get to know and interact with each other. To improve users' online experience and boost their levels of activities or engagement, many of these websites deploy recommender systems to help users find friends or users of interest.

Meanwhile, many social networks are based on reciprocity: a link between two users can only be established if both users agree to be connected (e.g., Facebook and LinkedIn). People in these reciprocal networks initiate or accept invitations to connect based on their own preferences or criteria. In addition, some of these reciprocal networks are also bipartite networks. A bipartite network has two types of nodes and an edge only exists between two nodes of different types (e.g., customer nodes and product nodes in a product purchase network; author and paper nodes in an authorship network). Examples of reciprocal and bipartite social networks include heterosexual dating networks (with male and female nodes), job application networks (with applicant and employer nodes), etc. The existence of reciprocity and bipartiteness in a social network poses new challenges to recommender systems.

In this research, we will address user recommendation in such reciprocal and bipartite social networks and use online dating network as a case study. In addition to being a typical bipartite social network with strong reciprocity, online dating is also very popular. 37% of all single American Internet users looking for a partner have visited an online dating website [1] and the online dating business is worth more than 2 billion British Pound [2]. Moreover, user/partner recommendation is especially important in online dating. While Facebook or LinkedIn users can directly search names of their acquaintance, users of online dating websites usually do not know many people in the social network and rely more on recommender systems to find potential partners. Thus improving user recommendation can potentially help an online dating website attract more users, improve their satisfaction, and generate more revenues.

**2. Related work**

In the literature, many approaches for friend/user recommendation in social networks utilize network structures. One popular approach is based on "friends of friends": the more common neighbors two users share, the more likely the two users will be connected [3]. Other approaches based on network structural features, such as triad [4] and multi-relational structures [5], have also been used for link prediction, including in reciprocal networks [6]. However, these approaches do not apply to bipartite social networks. For example, in a heterosexual dating network, only a male user can be connected to a female user. No matter how many common female neighbors two male users share, there will be no edge between the two male users.

Collaborative Filtering (CF) has been widely used to provide recommendations in bipartite networks, especially business transaction networks (e.g., the user-movie network in Netflix). However, links in a user-movie network are built in a unilateral way, because movies can only be passively chosen by users. By contrast, a reciprocal social network consists of autonomous individuals. A user's invitation to connect can be declined by another. In heterosexual dating, a successful match between a man and a woman depends on the reciprocal actions from both users. In other words, a dating partner recommender should consider a user's "taste" (who she/he likes), as well as "attractiveness" (who she/he will be liked by).

Little research has tackled the problem of user recommendation in bipartite reciprocal social networks. Researcher have leveraged sub-graph structures in a dating network and used data mining techniques for dating link prediction [7]. Reciprocal recommenders, such as CCR [8], mainly focused on users' "explicit profiles", including those that describe herself/himself and her/his preference of dating partners (e.g., age and education level). Then the bilateral matching of their profiles was used for recommendation.

In this research, we propose a CF recommender for reciprocal and bipartite social networks. We show that without using users' explicit profiles, the recommender can achieve very good performance, even better than the CCR model that uses both users' profiles and activities.

**3. Proposed approach**

In a user-movie network, movies picked by a user collectively reflect the user's taste. Similarly, looking at whom a user approaches in a dating network, we can infer the user's taste of dating partners. Meanwhile, attractiveness is also important for recommendations in reciprocal networks. For example, Mike (male) can certainly approach all the female users that he likes, but more importantly, he wants to get responses from these females. It is Mike's taste that affects whom he approaches through initial contacts, and his attractiveness that determines whether he can get responses. Considering the match of both taste and attractiveness between two users, our model tries to improve dating partner recommendations by boosting a user's chance of getting responses from potential partners that he/she likes.

**3.1. The context and notations**

We represent user activities in heterosexual online dating as a bipartite network, in which a user is a node and an edge in the dating network always connects a male and a female. In many dating websites, if user X is interested in user Y, she/he could approach Y by sending him/her a message, or an initial contact. If user Y is also interested in X, she/he could respond by sending a reply back to X, which constitutes a reciprocal contact between the two users[1]. Figure 1(a) shows an example of dating network.

Among all users $U$, we define users whom we provide recommendation to as service users $S$, where $S \subseteq U$. $N = |S|$ is the number of service users. $M = |U|$ is the total number of users ($N \leq M$). We separate service users from all users because CF-based models work better for users with more historical activities.

**3.2. The Baseline CF model**

We start with a very brief introduction to the baseline CF model, which will be used as one of the benchmarks. The classic CF model proceeds in 3 steps. The 1st step is to represent users' dating activities in a $N \times M$ contact matrix $C$. In the binary matrix $C$, $c_{i,j} = 1$ if user $i$ approached user $j$ through an initial contact, no matter whether user $j$ responded to user $i$ or not. Otherwise $c_{i,j} = 0$. Thus, a row in the matrix represents a service user's all activities in initiating contacts and could reflect his/her taste. Figure 1(b)-I shows an example of the contact matrix. The 2nd step is to quantify the similarity $s_{p,q}$ between service users $p$ and $q$. In this study, we use the cosine similarity between two row vectors for users $p$ and $q$ in matrix $C$. High similarity indicates that the two service users have similar tastes (i.e., they approached similar users). Users of opposite genders will have a similarity of zero because they do not approach any common users and their row vectors cannot have value 1 at the same column. The last step is to recommend dating partners. For a service user $p$, the model iterates through each target user $t$ ($t \in U$ and $t \neq p$) with whom user $p$ has not interacted with, and calculates a success score between $p$ and $t$ as $E_{p,t} = \sum_{k=1}^{N} s_{p,k} \times c_{k,t}$. The higher the $E_{p,t}$ score is, the more likely user $t$ will be rec-ommended to user *p*. The model basically suggests that if user *t* is approached by users whose tastes are similar to those of user *p*, then user *t* could be a potential partner for user *p*.

---

[1] While there are many ways to define what constitutes a reciprocal contact, many previous studies of dating network used the exchange of more than 1 message between two users as the standard.

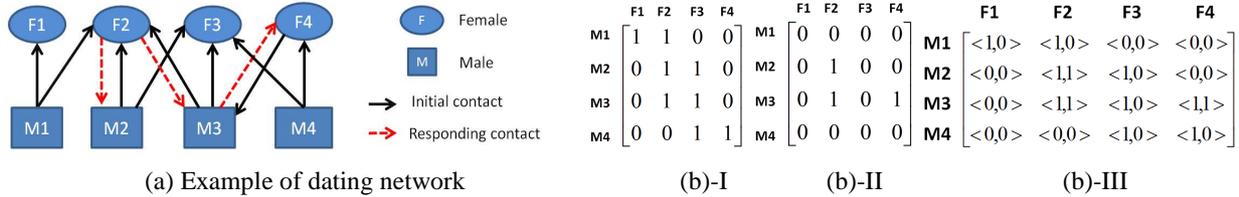

Figure 1. (a) An example bipartite dating network; (b) Corresponding contact matrices: (b)-I The baseline CF model, (b)-II The reciprocity-only model, (b)-III The Hybrid model, (Only male users are considered as service users).

### 3.3. The reciprocity-only model

The reciprocity-only model is proposed as another benchmark that only considers reciprocal contacts. It has a different contact matrix than the baseline CF model. In the reciprocity-only model's binary contact matrix $C$, $c_{i,j} = 1$ only if there is a reciprocal contact between users $i$ and $j$, and it does not matter who initiates the contact. Otherwise, $c_{i,j} = 0$. Even if users $i$ has approached $j$, as long as $j$ does not respond, $c_{i,j}$ will still have value 0. As a result, a row of this matrix represents both the taste and the attractiveness of a user. The calculation of similarity and the recommendation of partners are the same as the classic CF model. According to this model, user *t* could be recommended to user *p*, if *t* is interested in and attractive to users whose tastes and attractiveness are similar to user *p*.

### 3.4. The hybrid model

Although the reciprocity-only model can help to capture users' tastes and attractiveness, it has two limitations: (1) it ignores information on users' tastes and attractiveness when an initial contact is not responded. For the example in Figure 1(b)-II, M1 will have an empty row vector and we cannot track his taste; (2) it fails to leverage users' preference reflected by unresponded contacts. For instance, when F2 chose not to respond to M1's initial contact, it sug-gests M1's attractiveness does not match F2's taste. Then for users whose attractiveness is similar to M1, F2 may not be a good candidate. The two limitations also have implications for calculating the similarity between service users. In the same example, M3 and M4 have common taste and attractiveness (albeit in a discouraging way) because both approached F3 yet both failed to get responses. Such similarity between M3 and M4 is not captured by the reciprocity-only model.

Thus, we propose a hybrid model that considers both initial and reciprocal contacts. Formally, the contact matrix $C$ becomes a three dimensional matrix. For the purpose of simplicity, we still denote it with a two dimensional one, whose element $c_{i,j}$ is essentially a $1 \times 2$ vector $c_{i,j} =< c_{i,j,1}, c_{i,j,2} >$. If user $i$ has sent a message (initial or responding) to user $j$ (meaning $i$'s taste matches $j$'s attractiveness), then $c_{i,j,1} = 1$; otherwise, $c_{i,j,1} = 0$, meaning that $i$ is not interested in $j$. Similarly $c_{i,j,2} = 1$, if user $j$ is interested in user $i$; $c_{i,j,2} = 0$ otherwise. An example contact matrix for this hybrid model is in Figure 1(b)-III. This matrix can also be divided into two binary matrices, representing male users' taste toward female and female's taste towards male respectively, because female users' taste reflects male users' attractiveness and vice versa. However, for the purpose of simplicity in illustration, we decide to use a single contact matrix.

From such a contact matrix, we would like to take three types of similarity between service users into consideration: (1) taste--two users are interested in similar users; (2) attractiveness--two users attract similar users; and (3) unattractiveness--two users reject or are rejected by similar users. We denote the similarity between service users $p$ and $q$ as $s_{p,q} = \sum_{k=1}^{M} f(c_{p,k}, c_{q,k})$. When calculating inter-user similarity, users who share similar taste and attractiveness should have higher similarity scores than those with only similar taste (or attractiveness). Users whose taste and attractiveness are both different will have the lowest similarity. Thus we would like function $f$ to meet the following criteria: (1) $f(<x_1,y_1>,<x_2,y_2>)=f_{max}$ iff $x_1=x_2$ and $y_1=y_2$ and $x_1+x_2+y_1+y_2>0$; (2) $f(<x_1,y_1>,<x_2,y_2>)=f_{min}$ if $x_1 \neq x_2$ and $y_1 \neq y_2$; and (3) $f(<x_1,y_1>,<x_2,y_2>)=f_{min}$ if $x_1=x_2=y_1=y_2=0$. In our experiments, we pick $f(c_{p,k}, c_{q,k})$ as defined in Equation 1, where $\oplus$ represents exclusive OR. It yields three values: 2 for a match of both taste and attractiveness; 1 for a partial match of either taste or attractiveness; 0 for no match. $dgr(i)$ is the degree centrality of user $i$ in the un-directed and un-weighted dating network among users. Dividing by $dgr(p)+dgr(q)$ normalizes the similarity value. Alternative ways to calculate $f(c_{p,k}, c_{q,k})$ can also be defined. For instance, one can customize it so that $f(<1,0>, <1,0>) > f(<0,1>, <0,1>)$ if the match of taste is considered more important than the match of attractiveness in a given context.

$$f(<c_{p,k,1},c_{p,k,2}>,<c_{q,k,1},c_{q,k,2}>) = \frac{[\overline{(c_{p,k,1} \oplus c_{q,k,1})} + \overline{(c_{p,k,2} \oplus c_{q,k,2})}] * [(c_{p,k,1}+c_{p,k,2}) \cap (c_{q,k,1}+c_{q,k,2})]}{dgr(p)+dgr(q)} \quad \text{(Equation 1)}$$

To rank potential dating partners, we add a penalty factor when calculating the success score of recommending user $t$ to $p$: $E_{p,t} = \sum_{k=1}^{N} s_{p,k} \times g(c_{k,t})$. Function $g(c_{k,t})$, defined in (Equation 2, gives full weights to matches of both taste and attractiveness, while penalizes a partial match of either taste or attractiveness by $s$. It is also possible to assign different penalty factors for the match of taste and the match of attractiveness.

$$g(c_{k,t}) = \begin{cases} 1, & \text{if } c_{k,t} =<1,1>; \\ 1-s, (0<s<1) & \text{if } c_{k,t} =<1,0> \text{ or } <0,1>; \\ 0, & \text{otherwise} \end{cases} \quad \text{(Equation 2)}$$

In sum, the hybrid model extends the classic CF model in two ways. First, it considers both taste and attractiveness when calculating the similarity between service users. Users with similar taste and (un)attractiveness will have higher similarity scores than those who only share common taste or attractiveness. Second, it considers the match of both taste and attractiveness when recommending dating partners. Those who match both a service user's taste and attractiveness are more likely to be recommended than those who may only ignite unilateral interests.

## 4. Evaluation and discussions

The dataset used in this study is collected from a popular online dating website. It contains anonymized profiles and heterosexual dating activities of more than 47,000 users (60% being male) in two cities during a period of 196 days. There were 474,931 initial contacts, 79.8% of which were initiated by male users and 25.8% of them eventually became reciprocal contacts.

In our experiments, user activities in the first 98 days are used for training and the rest are for testing. We pick users who have sent 5 or more messages in both training and testing periods as service users, yielding a total of 6,628 service users (64% being male). The training set includes 159,558 pairs of contacts, and the testing set includes167,362 pairs.

We use two sets of metrics to evaluate recommenders' performance. The first is based on initial contacts (*IC*): IC Precision@K measures how many of the recommended *K* users were approached by the service user; IC Recall@K evaluates among all users approached by the service user, how many can be ranked within top *K*. The second set focuses on reciprocity--whether an initial contact is responded. Reciprocal-contact (*RC*) Precision@K evaluates how many of the recommended *K* users become a service user's reciprocal contacts; RC Recall@K measures among a service user's reciprocal contacts, how many can be ranked within top *K* by the recommender.

In addition to the baseline CF model and the reciprocity-only model, we also include two other CF recommenders in the comparison: CCR and Probabilistic Latent Semantic Analysis (pLSA). CCR [8] analyzes 7 manually selected user features (age, height, body type, education level, smoking, having children, marital status) from user profiles to calculate inter-user similarity and uses historical interactions of similar users to provide reciprocal recommendations. Originated as a text mining technique, pLSA has been adopted in CF-based recommendations [9]. It tries to find latent variables that represent the probabilistic affinity of observed two-mode data (e.g., topics in word-document data and preferences in buyer-product data). In this experiment, we vary the number of latent variables $v$ (7 different values from 5 to 200) and include in the comparison the one that yields the best performance ($v=10$).

**4.1. City-level evaluation results**

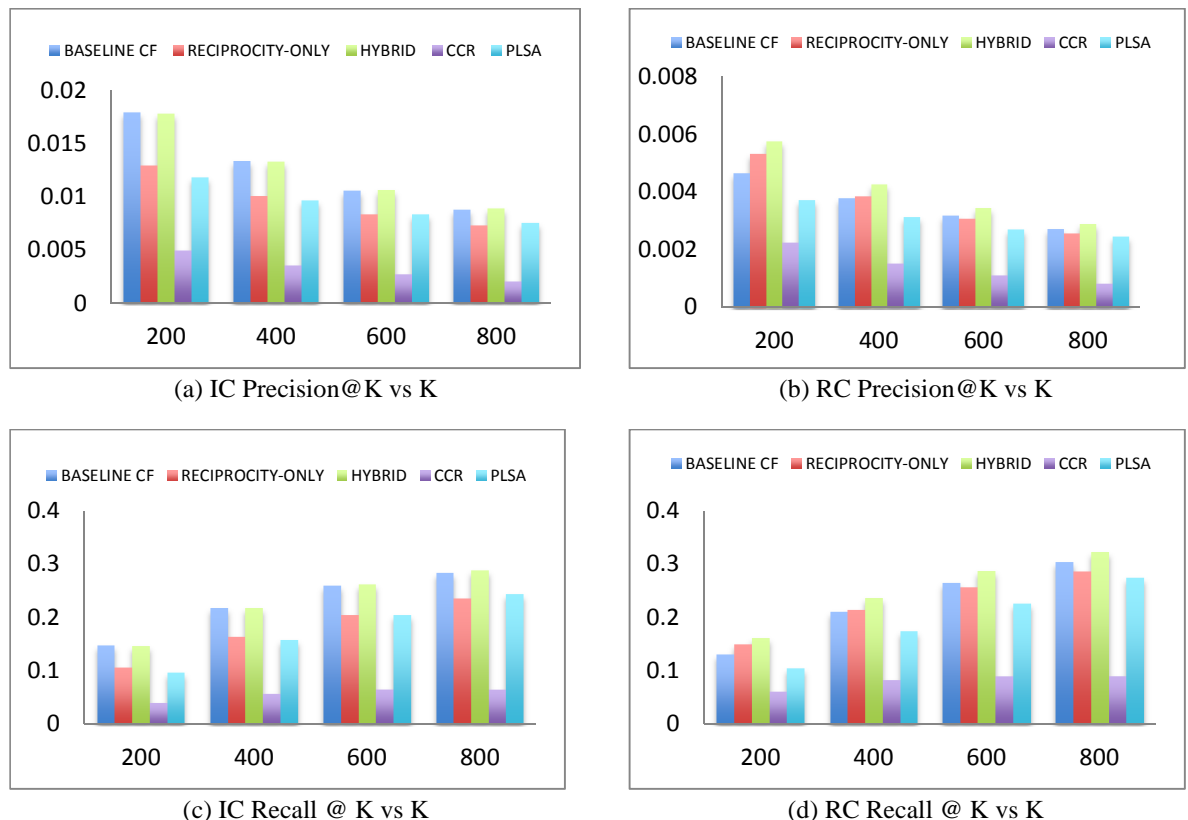

Figure 2. City-level comparison of 4 different models' performance for male users in a city.

At the city level, we aggregate all activities of service users, and examine recommenders'

performance. For the hybrid model, different penalty factor values have been tested. In general, when *s* increases, the hybrid model recommends more potentially reciprocal contacts, leading to better RC-based performance but slightly lower values for IC-based metrics. While all *s* values tested lead to decent performance of the hybrid model, we pick the one with *s=0.6* for the comparison as it has balanced performance on both IC and RC-based metrics.

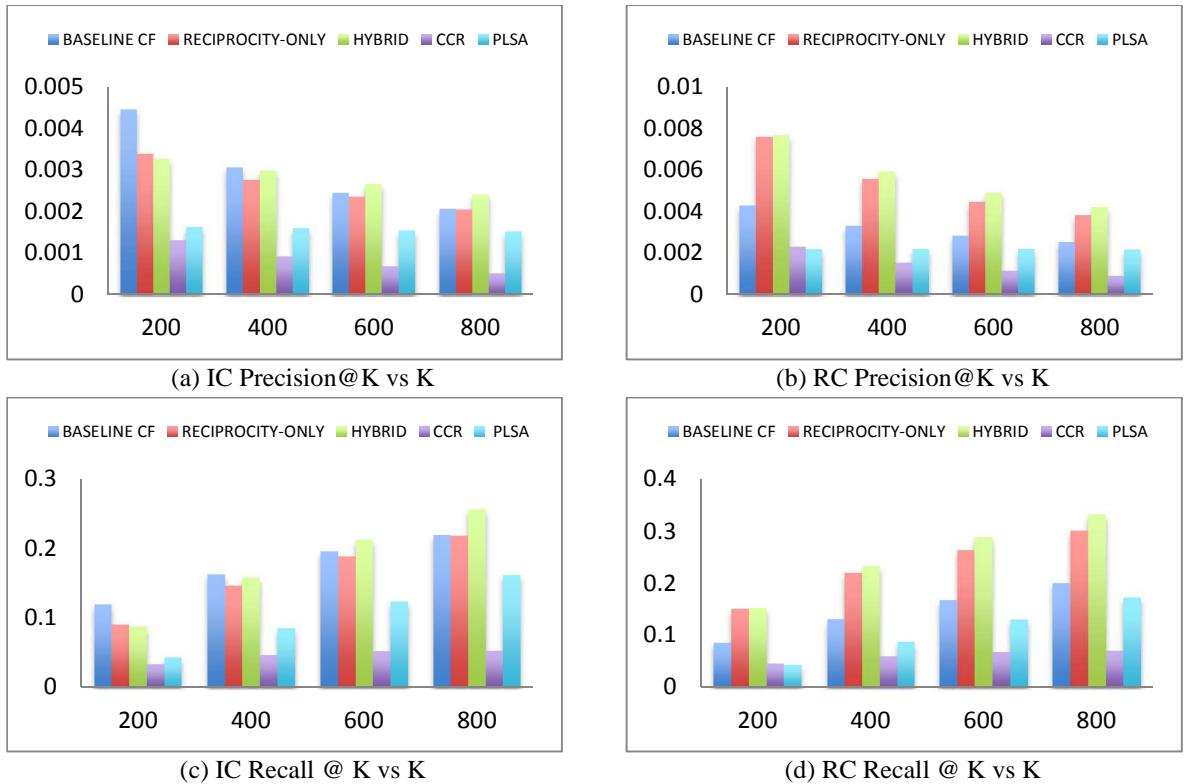

(a) IC Precision@K vs K
(b) RC Precision@K vs K
(c) IC Recall @ K vs K
(d) RC Recall @ K vs K

Figure 3. City-level comparison of 4 different models' performance for female users in a city.

Figure 2 and Figure 3 illustrate city-level performance of the five recommenders for male and female users respectively[2]. Overall, our hybrid model performs the best. On IC-based metrics, the hybrid model is similar with the baseline CF, especially for male users. For female users, it trails the baseline CF with lower *K* values, but surpasses the baseline CF when *K* increases.

---

[2] Due to the limit of space, we only show the results for one of the two cities in this paper, but the results from both cities are very similar.

More importantly, the hybrid model's RC-based metrics are the best among the five. Its advantages on RC-based metrics are more pronounced for female users. The reciprocity-only model performs worse than the baseline CF on IC-based metrics but its performance on RC-based metrics is a close second after the hybrid model. Although pLSA does better than CCR, both are still behind the other three recommenders.

## 4.2. Individual-level evaluation results

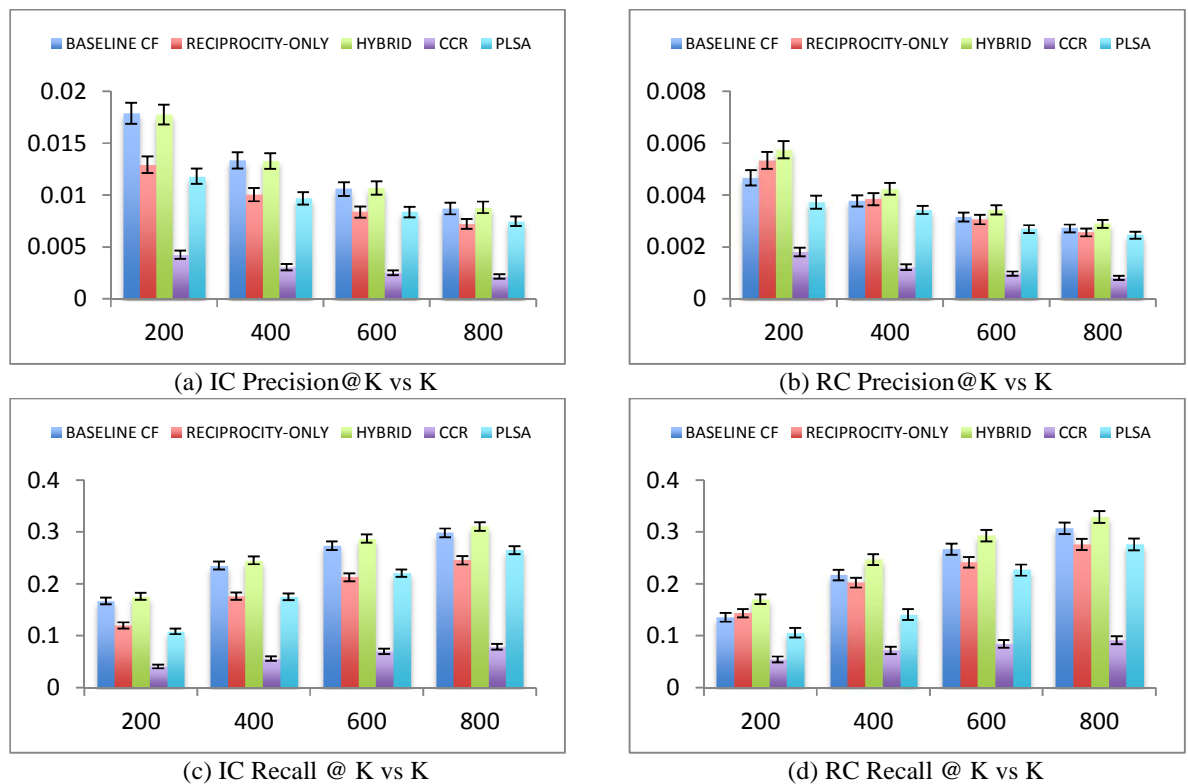

Figure 4. Individual-level comparison of 4 different models' performance for male users in a city. Vertical bars indicate 95% CIs.

Having been adopted by many studies, the macroscopic city-level evaluation is intuitive in illustrating a recommender's aggregated performance for a group of users. However, it can be biased when a recommender does a very good job for few service users with high activities in the testing period, but performs poorly for many less active users. For example, if a city has 3 ser-vice users—X, Y, and Z. X has 100 initial contacts in the testing period, and Y and Z have 10 each. Say a recommender (with *K=100*) can identify 70 of X's 100 initial contacts, 1 for Y's, and 1 for Z's. Although recommendations are hardly relevant for 2 out of the 3 service users, the recommender still achieves a decent city-level IC recall@100 = (70+1+1)/(100+10+10)=0.6.

To address this problem and bring in a different perspective that cares about each individual's experience with the recommender, we measure IC and RC-based metrics for each individual service user and compare the mean values of these individual-based metrics across a city for different recommenders. In the aforementioned example of 3 users, the average IC recall@100 would be (70/100+1/10+1/10)/3=0.24. Figure 4 and Figure 5 compare individual-level performance of the five recommenders. For this dataset, the results are similar to the city-level evaluation but help us understand the statistical significance of differences between models.

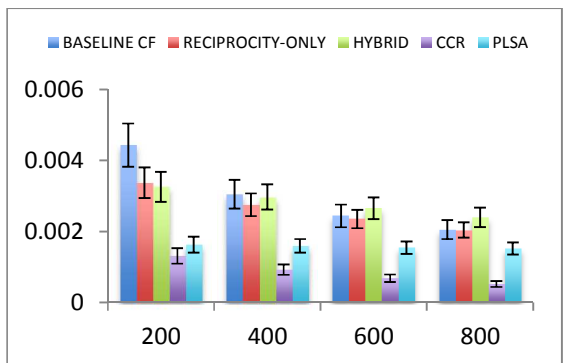
(a) IC Precision@K vs K

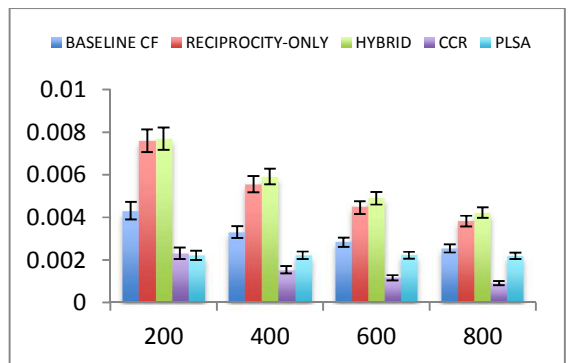
(b) RC Precision@K vs K

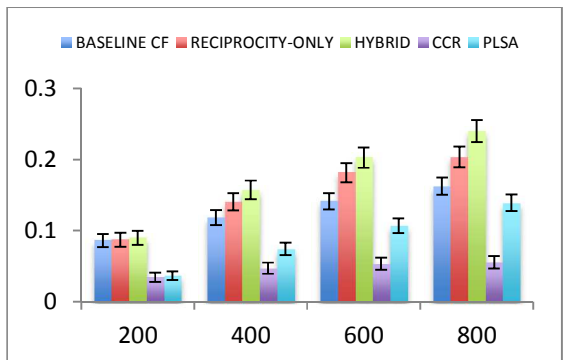
(c) IC Recall @ K vs K

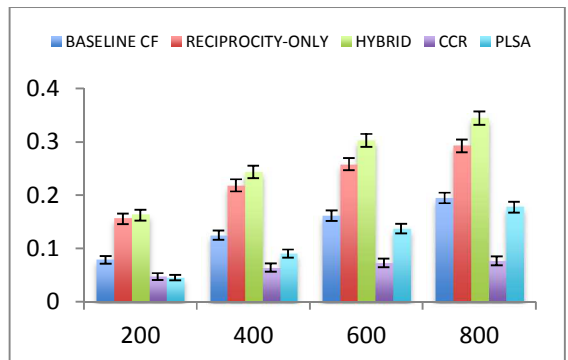
(d) RC Recall @ K vs K

Figure 5. Individual-level comparison of 4 different models' performance for female users in a city. Vertical bars indicate 95% CIs.

### 4.3. Discussions

Our experiments show that the hybrid model can recommend partners who are attractive to and interested in a service user. Specifically, if a service user approaches a partner recommended by the hybrid model, he/she will have a better chance of getting responses. In addition, this performance is achieved without compromising the ability to recommend partners that a service user likes (measured by IC-based metrics). By contrast, partners recommended by the baseline CF model are less likely to be interested in the service user. The reciprocity-only model's performance is generally in between the hybrid and the baseline CF models.

The three model's performance on IC and RC-based metrics can be explained by how they leverage different types of information about users' taste and attractiveness. The baseline CF model uses all information on users' taste but discards information on attractiveness. Thus it has slightly better IC-based performance, which relies only on the accurate capture of taste, but lags far behind the hybrid and the reciprocity-only models on RC-based metrics. The reciprocity-only model considers attractiveness (reflected by replies to initial contacts), but discards information about users' taste and unattractiveness reflected in non-reciprocal contacts. As a result, it does poorly in recommending partners that match a service user's taste, but improves the chance that a recommended partner is attracted by the service user. The hybrid model mixes three types of information--one's taste, attractiveness, and unattractiveness, and has the best overall performance.

To illustrate the performance of the hybrid model on different types of users, we separate service users in our experiments into two groups: the "successful recommendation" (SR) group includes service users to whom the hybrid model can successfully recommend at least one reciprocal contact. Other service users are in the "unsuccessful recommendation" (UR) group. According to t-test results, both male and female users in the SR group have sent more messages than their peers in the UR group (85.6 vs 51.9 messages per male user, and 70.2 vs 60.0 messages per female user)[3]. This is also the characteristic of CF recommenders--the more actively a user approaches others, the more information a recommender gets about his/her taste and attractiveness, and the more effective recommendations become. We also find that all models perform better for male on IC-based metrics, but female users still enjoy similar RC-based performance with male. It turns out female users on average (63.0 messages per user) send fewer messages than male (68.2 per user)[4], which means less information is available about their taste, leading to lower IC-based metrics. However, when a female user does approach others, she has a better chance of getting responses (41.7% vs male's 21.4%). Thus the model can still capture their attractiveness through these reciprocal contacts and achieve RC-based performance similar to male.

Besides the level of user activities, we also compare users' attributes in their online profiles. For male and female in the SR and UR groups, we first find the average distributions of each of their 14 personal attributes (e.g., race and height). Then for each attribute and each gender, we calculate the Euclidean distance between the average distribution vectors for users in the two groups. After ranking the attributes on descending distances, "body type" and "number of photos" emerge as the top differentiating attributes for male users between the two groups; and "children" and "body type" for female. Specifically, the hybrid model performs better for male with "athletic" body type, female with "athletic" or "fit" body type; but worse for male with "average" body and female with "average" or "thin" body type. Meanwhile, female in the SR group tend to "want many kids", and more female users in the UR group "want no kid". Also, male users in the SR group uploaded more photos than those in the UR group.

---

[3] $P\text{-}value<0.05$.
[4] $P\text{-}value<0.05$.

Last but not least, the performance of CCR model is generally the worst even though it also considers reciprocity. We conjecture that this is because our dataset contains different personal attributes, and different ways to organize attributes (e.g., numeric or categorical, how many categories for categorical data), with the dataset used for CCR. In fact, the performance of CCR has highlighted one advantage of our hybrid CF recommender--it does not need to deal with personal attributes, which can often be inconsistent across different dating services and inaccurate [10].

## 5. Conclusions and future work

In this research, we propose a new CF model to improve user recommendation in bipartite and reciprocal social networks. Leveraging users' historical activities in approaching others and getting responses (or not), the model focuses on users' taste and attractiveness in establishing bilateral connections. Using an online dating network as a case study, we illustrate that the new model performs well on recommending both unilateral and reciprocal contacts. In other words, the new model can better recommend partners that matches a user's taste and attractiveness. We also analyze the characteristics of users that can get effective recommendations from the proposed model. The outcome of this research can also be incorporated into existing recommenders for other reciprocal bipartite networks, such as college admission network (high school students and universities as nodes), job-hunting network (applicants and employers as nodes), and so on.

There are also many possible directions for future research. We would like to further validate the effectiveness and robustness of the hybrid model through sensitivity analysis (e.g., change the pool of users, the testing/training periods, etc). Also, by carefully incorporating the profile information of the users, we also hope to find a method that can help to solve the cold-start problem for CF-based recommenders.

**Acknowledgement**

Kang Zhao was supported by the University of Iowa Old Gold Summer Fellowship. Mo Yu was partially supported by the Pennsylvania State University and the National Science Foundation under IGERT award # DEG-1144860.

**Author bio**

**Kang Zhao** is an Assistant Professor at Department of Management Sciences, The University of Iowa. His research interests include social computing and business analytics. He obtained his Ph.D. from Penn State University.

**Xi Wang** is a Ph.D. student of Interdisciplinary Graduate Program in Informatics at The University of Iowa. Her research interests include recommendation and social support in online social networks, and association rules mining. She received her M.S. from Beijing University of Posts and Telecommunications.

**Mo Yu** is a Ph.D. student in Information Sciences and Technology, Penn State University, and a research associate in Big Data Social Science IGERT program. His research interests are data mining and social network analyses. He got his bachelor's degree from Peking University.

**Bo Gao** is a software engineer at Network Center, Beijing Jiaotong University, and a Ph.D. student at the State Key Laboratory of Rail Traffic Control and Safety. His research interests include transportation networks and safety, social networks, and scientific computing.

**Author Contact Information:**

**Kang Zhao**
S224 PBB,
Iowa City, IA 52242, USA
Phone: 1-319-335-3831   Fax: 1-319-335-0297   Email: kang-zhao@uiowa.edu

**Xi Wang**
411 Emerald St  E18
Iowa City, IA 52246, USA
Phone: 1-319-855-3162   Email:xi-wang-1@uiowa.edu

**Mo Yu**
324 IST Building,
University Park, PA 16802, USA
Phone: 1-814-865-6178   Email: muy145@psu.edu

**Bo Gao**
Beijing Jiaotong University Network Center
Beijing 100044, China.
Phone: 86-10-51685081  Email: gaobo@bjtu.edu.cn